\documentclass[useAMS,usenatbib]{mn2e}
\usepackage{graphicx}
\usepackage{epsfig}
\newcommand{\LA}{\mbox{\raisebox{-0.6ex}{$\stackrel{\textstyle<}{\sim}$}}}
\newcommand{\GA}{\mbox{\raisebox{-0.6ex}{$\stackrel{\textstyle>}{\sim}$}}}
\title{Dynamic Boundaries of Event Horizon Magnetospheres}
\author[Brian Punsly]{Brian Punsly \\
4014 Emerald Street No.116, Torrance CA, USA 90503 and \\
International Center for Relativistic Astrophysics,
I.C.R.A.,University of Rome La Sapienza, I-00185 Roma, Italy\\
E-mail: brian.m.punsly@L-3com.com or brian.punsly@gte.net}
\begin{document}
\maketitle \label{firstpage}
\begin{abstract}This Letter analyzes 3-dimensional simulations
of Kerr black hole magnetospheres that obey the general relativistic
equations of perfect magnetohydrodynamics (MHD). Particular emphasis
is on the event horizon magnetosphere (EHM) which is defined as the
the large scale poloidal magnetic flux that threads the event
horizon of a black hole (This is distinct from the poloidal magnetic
flux that threads the equatorial plane of the ergosphere, which
forms the ergospheric disk magnetosphere). Standard MHD theoretical
treatments of Poynting jets in the EHM are predicated on the
assumption that the plasma comprising the boundaries of the EHM
plays no role in producing the Poynting flux. The energy flux is
electrodynamic in origin and it is essentially conserved from the
horizon to infinity, this is known as the Blandford-Znajek (B-Z)
mechanism. To the contrary, within the 3-D simulations, the lateral
boundaries are strong pistons for MHD waves and actually inject
prodigious quantities of Poynting flux into the EHM. At high black
hole spin rates, strong sources of Poynting flux adjacent to the EHM
from the ergospheric disk will actually diffuse to higher latitudes
and swamp any putative B-Z effects. This is in contrast to lower
spin rates, which are characterized by much lower output powers and
modest amounts of Poynting flux are injected into the EHM from the
accretion disk corona.
\end{abstract}

The long term 3-D simulations discussed in
\cite{dev03,dev05,hir04,kro05,haw06} (\textbf{HK}, hereafter) offer
an important virtual laboratory for studying the physics of black
hole driven jets. These simulations readily evolve to a
configuration with a net accreted poloidal magnetic flux that is
trapped within the accretion vortex or funnel. This region is the
black hole magnetosphere and it supports a jet dominated by
electromagnetic energy for rapidly rotating black holes. In a
previous paper, \cite{pun07}, it was shown that the physics
conducive to an ergospheric disk (first described in \cite{pun90})
existed in the high spin simulation, $a/M=0.99$ (where the black
hole mass, $M$, and the angular momentum per unit mass, $a$, are in
geometrized units), known as KDJ. The ergospheric disk jet (EDJ) is
launched from the plasma near the equatorial plane of the
ergosphere. The event horizon magnetosphere (EHM) is comprised of
poloidal flux that threads the event horizon (see the middle frame
of figure 1). This paper explores the sources of Poynting flux in
the EHM in the context of the 3-D simulations. In the EHM, there are
four possible sources of Poynting flux that follow from energy
conservation.
\begin{enumerate}
\item Energy can be transferred from the plasma to the
electromagnetic field. However, the EHM is virtually evacuated of
plasma due to the centrifugal barrier and there is little energy to
transfer to the field in perfect MHD.
\item Any other sources must be surface terms at the boundary of the EHM.
The first possibility is the electrodynamic component that is
associated with the event horizon boundary surface. This is
customarily called the Blandford-Znajek (B-Z) effect \cite{blz77}.
\item The second boundary surface, which is associated with the EDJ, is the equatorial accretion flow in the ergosphere.
\item Outside the ergosphere, the EHM is bounded by the
accretion disk and the disk corona. A source on this boundary will
be called a coronal piston (see figure 2).
\end{enumerate}
\par It has been typically assumed that a Poynting jet in the EHM is
confirmation of a B-Z process at work. However, it will be shown
that the two simulations, KDJ with $a/M=0.99$ and KDH with
$a/M=0.95$, indicate that this can be a misleading interpretation in
general. The energy output of KDJ is dominated by the EDJ
\cite{pun07}. KDH provides an interesting contrast to KDJ, because
the EDJ is not nearly as powerful as it is in KDJ in the late time
data slices. In KDJ, the Poynting flux in the EHM at large distances
from the hole, $r\sim 100 M$, is dominated by the energy flux from
the EDJ that is gradually diffusing to higher latitudes within the
funnel as it propagates outward. For KDH, the coronal piston that
was discussed in \textbf{HK} in the context of driving the enormous
mass flux in the "funnel wall jet" also injects significant amounts
of Poynting flux into the funnel as a second order effect.
\begin{figure}
\vspace{-1.5cm}
 \hspace{-0.90cm}
\includegraphics[width=98 mm]{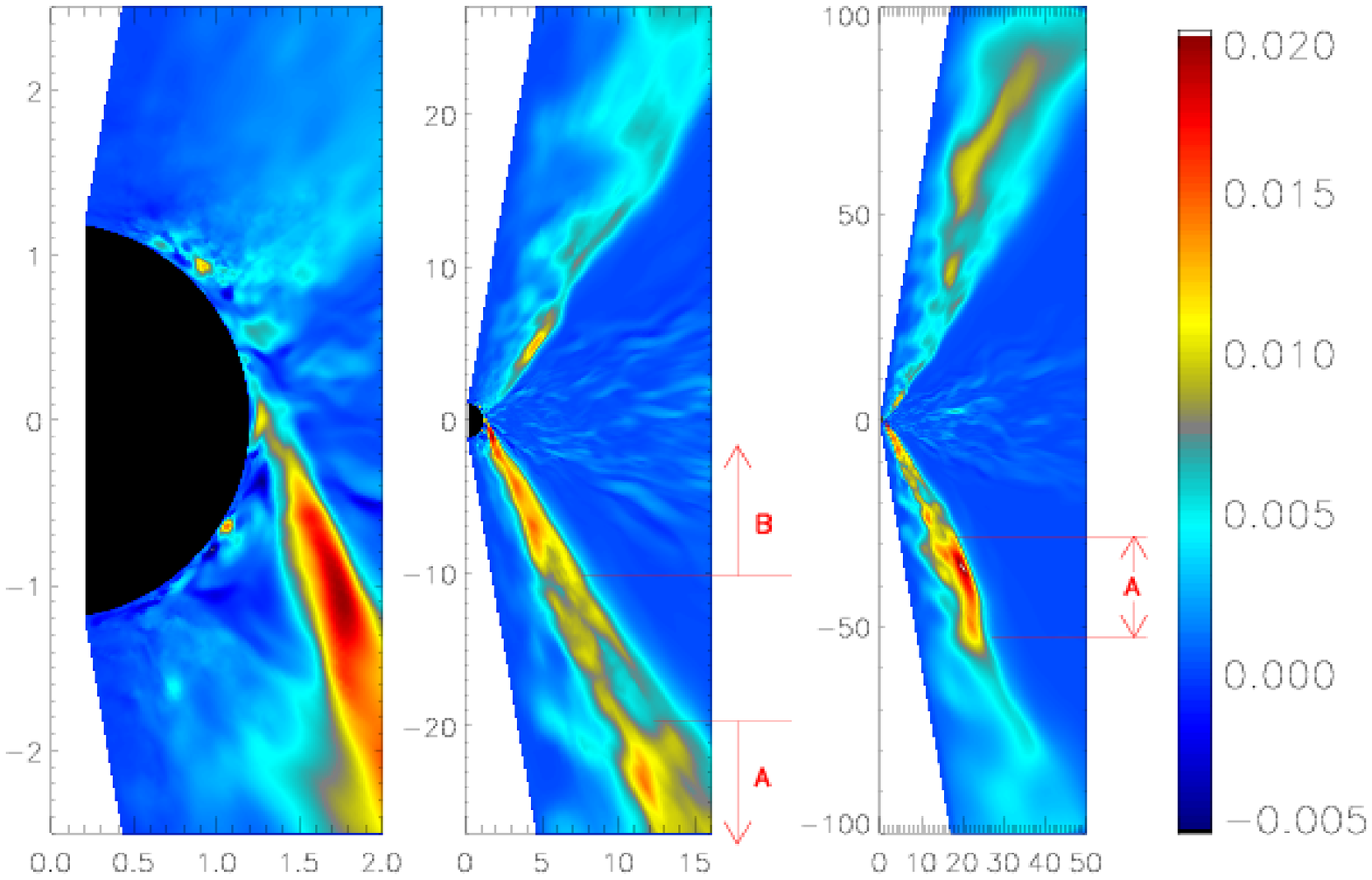}
\end{figure}
\begin{figure}
\vspace{-2.3cm}
 \hspace{-0.90cm}
\includegraphics[width=98 mm]{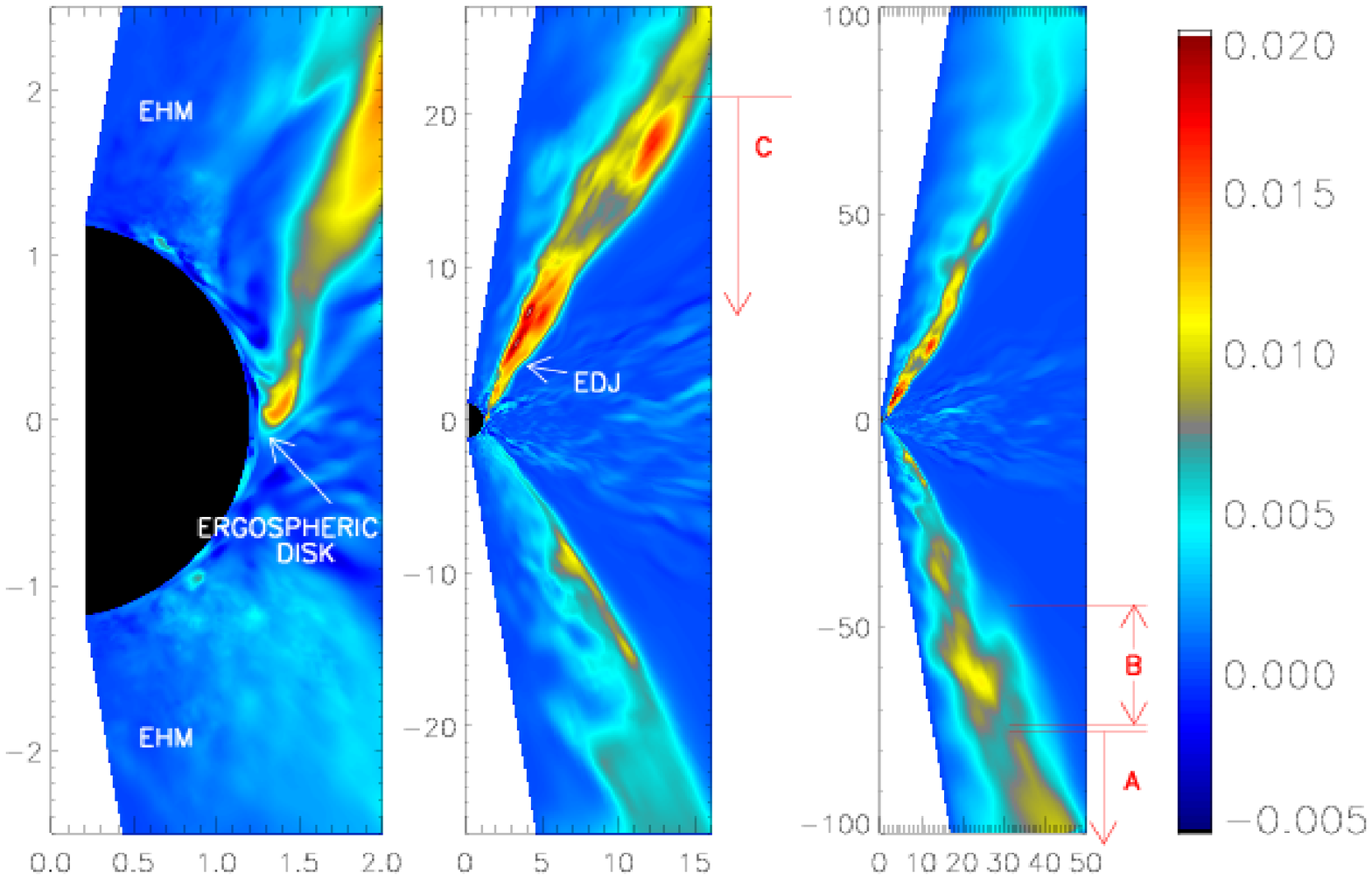}
\vspace{-0.7cm}
 \hspace{-0.90cm}
\includegraphics[width=98 mm]{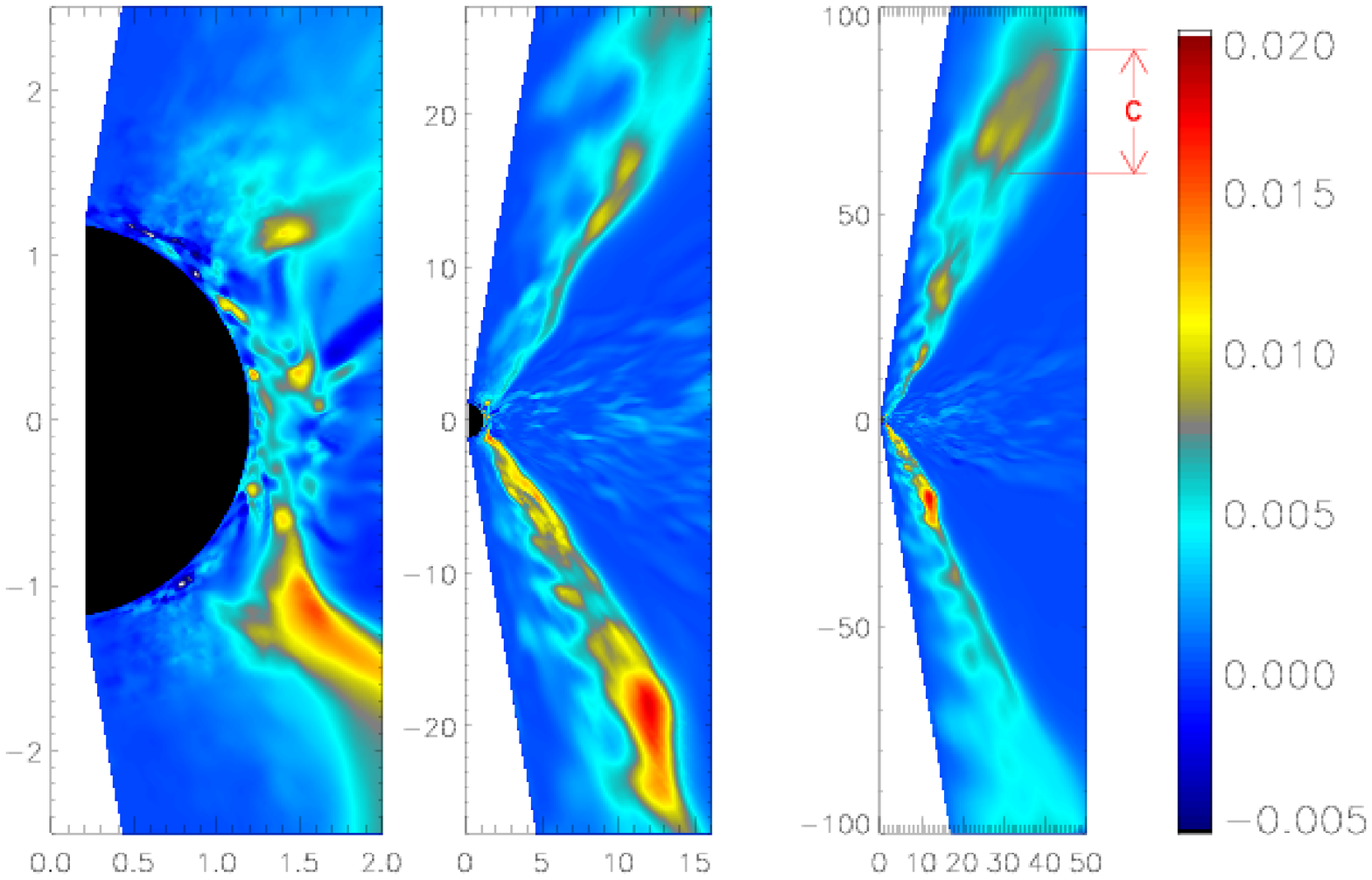}
 \caption{The Poynting flux in KDJ. The color bar is in code units. The rows are in chronological order, t= 9840 M (top),
 t=9920 M (middle) and t= 10000 M (bottom). There is no data clipping, saturated regions are white. The region
 $r < r_{in}=1.203 M$ is black and $r_{+} = 1.141 M $.}
\end{figure}
\section{The 3-D Simulations}J. Krolik and J. Hawley have
generously shared the data for the last three time slices of KDH and
KDJ, at t = 9840 M, t= 9920 M and t=10000 M. The simulations are
performed in the Kerr metric (that of a rotating, uncharged black
hole), $g_{\mu\nu}$. Calculations are carried out in Boyer-Lindquist
coordinates $(r,\theta,\phi,t)$. The reader should refer to
\textbf{HK} and references therein for details of the simulations.
We only give a brief overview of the most relevant details. The
initial state is a torus of gas in equilibrium that is threaded by
concentric loops of weak magnetic flux that foliate the surfaces of
constant pressure. The magnetic loops are twisted azimuthally by the
differentially rotating gas. This creates significant magnetic
stress that transfers angular momentum outward in the gas,
initiating a strong inflow that is permeated by magneto-rotational
instabilities (MRI). The end result is that after t = a few hundred
M, accreted poloidal magnetic flux gets trapped in the accretion
vortex or funnel (with an opening angle of $\sim 60^{\circ}$ at the
horizon tapering to $ \sim 30^{\circ} - 35^{\circ}$ at $r > 20 M $).
This region is the black hole magnetosphere and it supports Poynting
flux. The surrounding accretion flow is very turbulent. The strong
transients die off by t = 2000 M, so the late time data dumps are
the most physically relevant \cite{haw06}.
\par Numerically, the problem is formulated on a grid that is 192 x
192 x 64, spanning $ r_{in} < r < 120M $, $8.1^{\circ}< \theta <
171.9^{\circ}$ and $0 < \phi < 90^{\circ}$. The inner calculational
boundary, $r_{in}$, is located close to, but just outside of the
event horizon, $r_{+}$, where the coordinates are singular. The
$\phi$ boundary condition is periodic and the $\theta$ boundary
conditions are reflective. Zero-gradient boundary conditions are
employed on the radial boundaries, where the contents of the active
zones are copied into the neighboring ghost zones. MHD waves
propagate slower than the speed of light, therefore the
gravitational redshift creates a magneto-sonic critical surface
outside of $r_{+}$ from which no MHD wave can traverse in the
outward direction \cite{pun01}. The philosophy was to choose
$r_{in}$ to lie inside the magneto-sonic critical surface, thereby
isolating it from the calculational grid. There are also steep
gradients in the metric derived quantities as $r_{+}$ is approached.
This is handled by increasing the resolution of the grid near
$r_{in}$ with a cosh distribution of radial nodes. The validity of
the numerics of this method was verified, near $r_{in}$, in
\cite{dev01} by comparing simulations to simple analytic solutions.
Even so, the simulations are closely monitored to look for unnatural
boundary reflections. We also note the 3-D simulations in
Kerr-Schild coordinates (which are nonsingular on the horizon) in
\citet{fra07}. To test the code, they ran simulations of magnetized
tori that were initiated from identical input parameters to those
used by Hawley et al. In the words of C. Fragile (private
communication), the results were "remarkably similar." Even though
this was only verified for a/M=0.9, it is compelling. Consequently,
for the purposes of this study it was concluded that the numerics
were reliable inside the ergosphere.
\section{The KDJ Simulation}  The conservation
of global, redshifted, energy flux, defined in terms of the
stress-energy tensor, is simply, $\partial(\sqrt{-g}\, T_{t}^{\,
\nu})/\partial(x^{\nu})=0 $ \citep{pun07}. The four-momentum
$-T_{t}^{\, \nu}$ has two components: one from the fluid,
$-(T_{t}^{\, \nu})_{\mathrm{fluid}}$, and one from the
electromagnetic field, $-(T_{t}^{\, \nu})_{\mathrm{EM}}$. The
quantity $g = -(r^{2} + a^{2} \cos^{2}{\theta})^{2}\sin^{2}{\theta}$
is the determinant of the metric. The integral form of the
conservation law arises from the trivial integration of the partial
differential expression \cite{thp86}. It follows that the poloidal
components of the redshifted Poynting flux are $S^{\theta}=
-\sqrt{-g}\, (T_{t}^{\, \theta})_{\mathrm{EM}}$ and $S^{r}=
-\sqrt{-g}\, (T_{t}^{\, r})_{\mathrm{EM}}$. We can use these simple
expressions to understand the EHM Poynting jet in KDJ. Figure 1 is a
plot of $S^{r}$ in KDJ viewed at three different levels of
magnification, at t= 9840 M (top row), t= 9920 M (middle row) and
the bottom frames are at t = 10000 M. Each frame is the average over
azimuth of each time step. This greatly reduces the fluctuations as
the accretion vortex is a cauldron of strong MHD waves. The
individual $\phi=\mathrm{constant}$ slices show the same dominant
behavior, however it is embedded in large MHD fluctuations. The left
hand column shows strong beams of $S^{r}$ coming from near the black
hole. In \citet{pun07}, it was shown that the source of these beams
was $S^{\theta}$ that was created by the ergospheric disk. The base
of the EDJ emerging from the dense equatorial plasma is well
resolved in both $r$ and $\theta$ with $\approx 20$ and $>30$ grid
zones, respectively.
\par In this paper, we turn our attention to the
propagation of individual flares from the ergospheric disk out to
the outer calculational boundary at $r = 120 M$. Even though the
time sampling is very coarse in the data dumps ($\Delta t = 80 M $),
we can understand the propagation of the EDJ because of the wide
angle views available in the right hand column of figure 1. We track
the EDJ evolution by identifying the strong knots or flares in
figure 1 based on the following reasoning. As discussed in
\cite{pun01}, the $S^{r}$ flares will propagate at the speed of an
MHD discontinuity as modified by the plasma bulk flow velocity. The
plasma near the edge of the vortex has accelerated to $v_{r}> 0.9c$
by r= 30 M. So the flares of $S^{r}$ should propagate radially at
$V_{flare}\, \LA \, c $ for $r>30M$. Without having the benefit of
the detailed time evolution, this upper bound is the best estimate
that we can make for $V_{flare}$. First, consider the strong knot,
"C," at t=10000 M in the right hand frame. Label the outer radial
extent of knot C at t = 10000 M by $r_{+C}(t = 10000 M)= 100.9 M$
and inner radial edge by $r_{-C}(t = 10000 M)= 65.9 M$. Translating
this MHD discontinuity back in time to t = 9920 M is equivalent to a
radial displacement $V_{flare}\Delta t = V_{flare} (-80 M/c)\, \GA
\, -80 M $. Thus at t= 9920M, knot "C" should extend from the
ergospheric disk to $r_{+C}(t = 9920 M) \, \GA \,  20.9 M $. This is
verified by the red patches in the middle frame at t=9920 M.
\par Next consider the strong knot, "A," at t= 9840 M in the right hand frame. Label the outer radial
extent of knot A at t = 9840 M by, $r_{+A}(t = 9840 M)= 58.9 M$ and
inner radial edge by $r_{-A}(t = 9840 M)= 22.0 M$. Time translating
this feature to t = 9920 M implies that $r_{-A}(t = 9920 M)\, \LA \,
102.0 M$, so it must be visible near the edge of the right hand
frame at t = 9920 M. Furthermore, unless the flare is propagating
inordinately slowly, $V_{flare} < 0.75 c $, $r_{+A}(t = 9920 M)$
will be beyond the outer boundary of the plot. There is only one
plausible feature at t =9920 M. Finally, there is a strong flare "B"
that is emerging from the ergospheric disk at t = 9840 M in the left
and middle frames. Thus, at t = 9920 M, some portion of the flare
must be within 80 M of the black hole, hence the identification of
"B" in the right hand frame.
\par Figure 1 demonstrates a type (iii) source in the notation of the introduction. At $r \approx
1.5 M - 2.0 M$, the EDJ enters the EHM from the periphery (left hand
frames). The EDJ gets quickly linked into the EHM because the
ergopsheric disk magnetosphere in KDJ is comprised of small patches
of twisted vertical flux that become intertwined with the
large-scale flux in the EHM on scales of $\sim$ 1M - 2M
\citep{pun07}. After this rapid injection, the $S^{r}$ in the EDJ
keeps spreading towards the pole as it propagates outward. At the
time steps that were made available to this author, the EDJ is the
predominant source of $S^{r}$ in the EHM. By $r\approx 100 M$, the
EDJ is flooding the EHM, even close to the polar axis (it should be
noted that the total energy flux is larger than $S^{r}$,
$\approx20\%$ is in mechanical form in the EHM at $r\approx 100 M$).
The relevance to this discussion is that the EHM is inundated with
$S^{r}$ that was not created on field lines that thread the horizon,
but on flux entrapped within the equatorial accreting plasma. The
slow diffusion of $S^{r}$ poleward at $r > 30 M$ is most likely
regulated by numerical diffusion. This might seem like a problem
from a numerical point of view, but physically this is not nearly as
much of a concern from a qualitative standpoint. Perfect MHD is just
a simple tractable method of dealing with the plasma physics. A
realistic, high temperature, jet plasma is likely to have anomalous
resistivity from a variety of sources, \cite{som00,tre01}, and the
diffusion of field energy should naturally occur. The simulation
cannot accurately describe the diffusion rate. However,
qualitatively speaking it indicates that if the jet propagates
extremely far from the hole ($r \gg 120M$), regardless of the exact
details of the diffusion microphysics, the EDJ energy flux is likely
to get smeared out towards the polar region.
\begin{figure*}
\hspace{-0.75cm}
\includegraphics[width=85 mm]{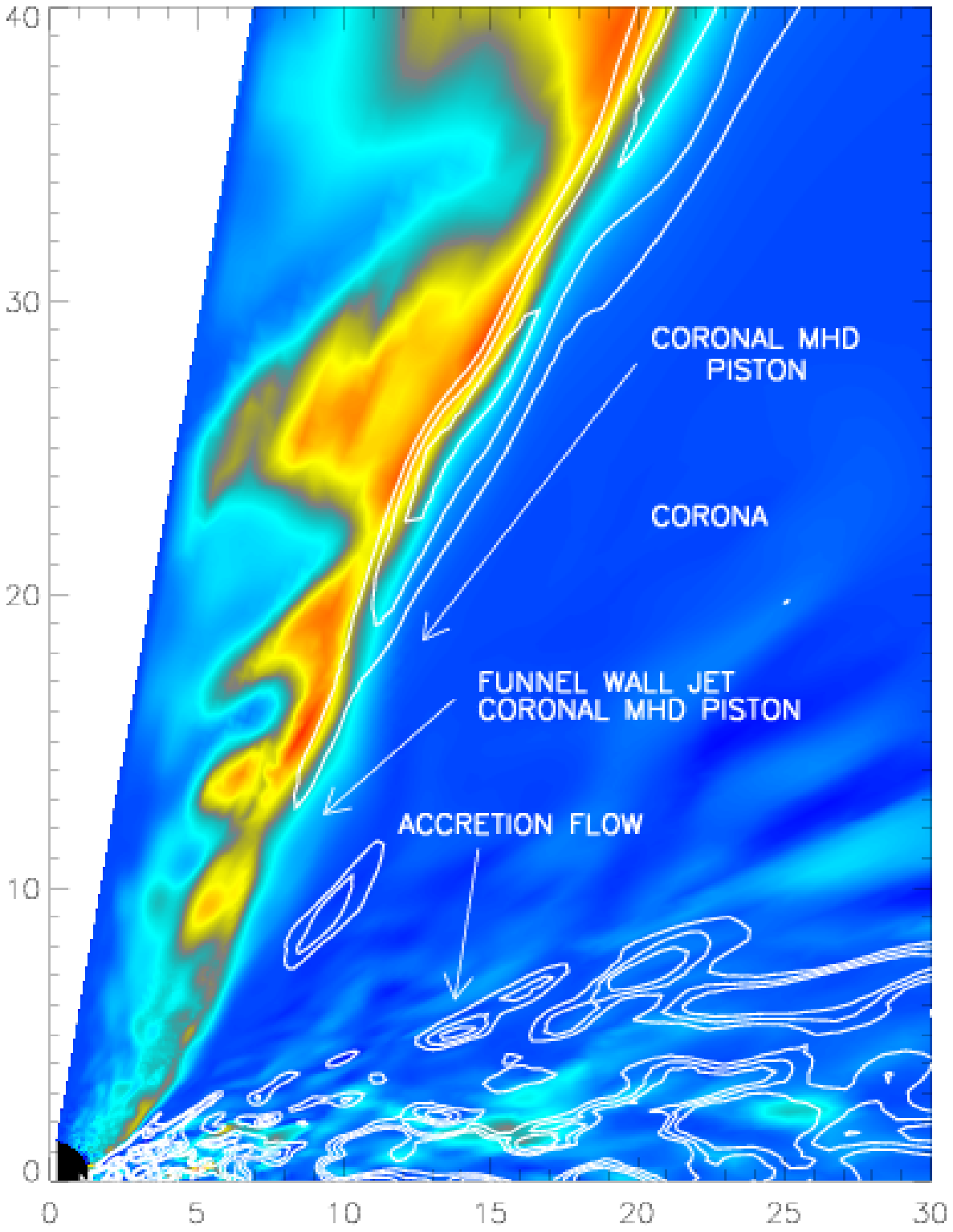}
\hspace{-3.8cm}
\includegraphics[width=85 mm]{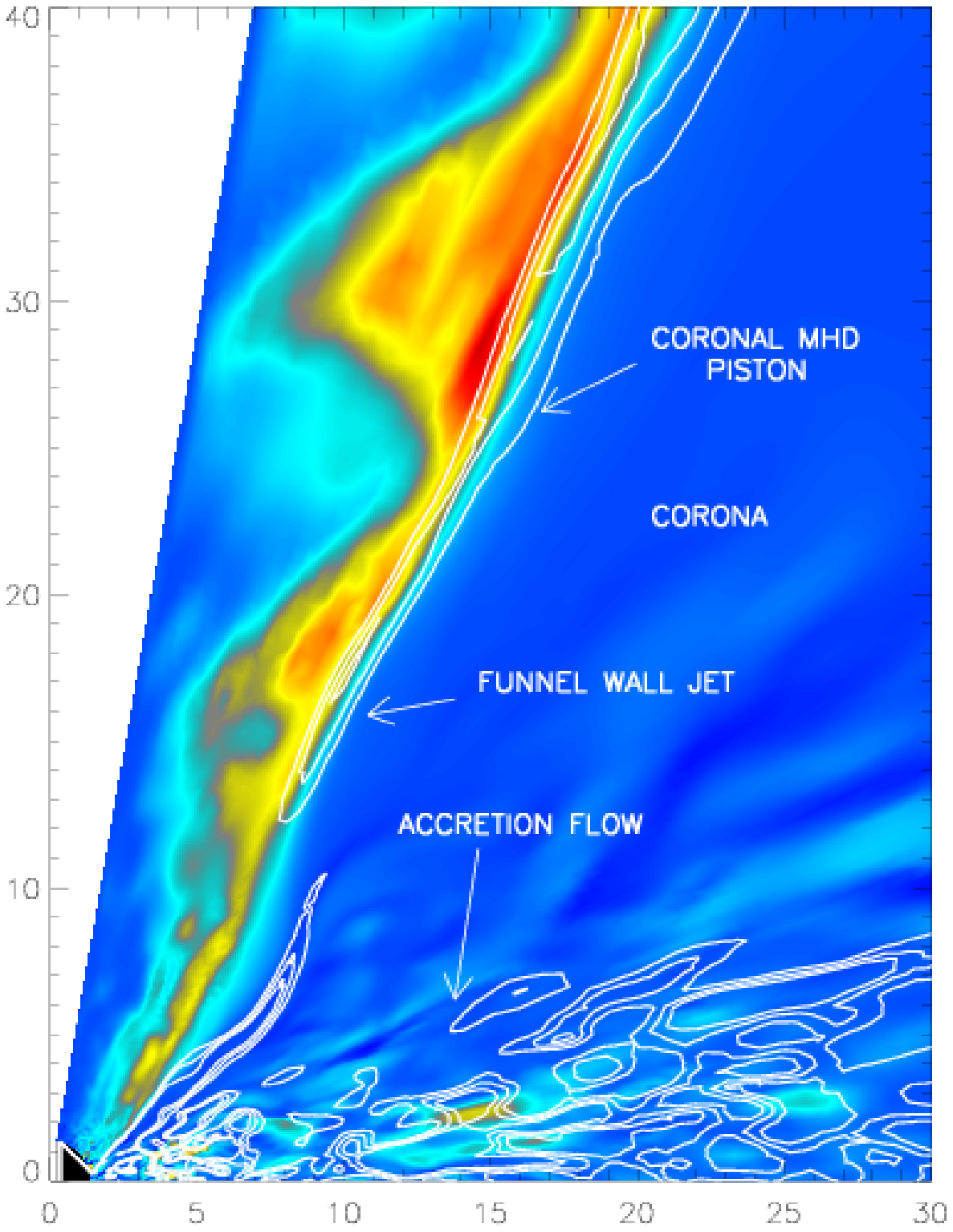}
\hspace{-3.8cm}
\includegraphics[width=85 mm]{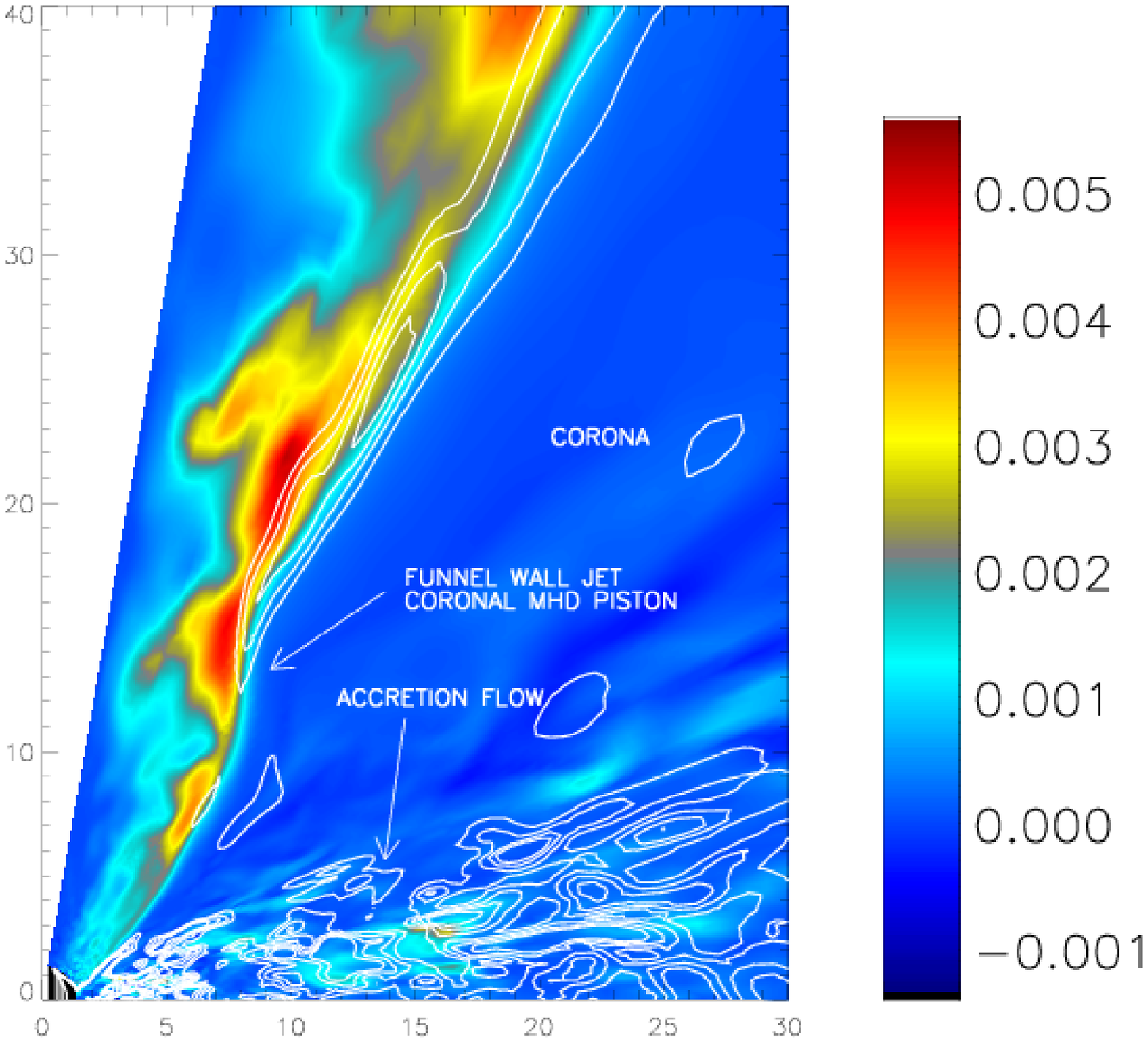}
 \caption{The coronal MHD piston is illustrated by these plots of the Poynting flux in KDH. The color bar is in code units. The frames are in chronological order, t= 9840 M (left),
 t=9920 M (center) and t= 10000 M (right). There is no data clipping, saturated regions are white. The overlayed white contours represent, $P_{r}$, the radial momentum
 flux described in the text. At each coronal piston
location there is a large pressure flare, see figure 3. Inside of
$r_{in}=1.403 M$ ($r_{+} =1.312M$) is black.}
\end{figure*}
\section{The KDH Simulation} There is a relatively weak EDJ in KDH that is noticeable at t = 9840 M, but it is otherwise negligible. This
circumstance allows for the detection of weaker sources of $S^{r}$
that would otherwise be swamped by a strong EDJ. Figure 2 is a plot
of $S^{r}$ for these three time slices in chronological order, going
from left to right. Each frame is the average over azimuth of the
time step. The contours of the radial momentum flux due to mass
motion, $P_{r}\equiv \sqrt{-g}\rho U^{r}U_{r}$ (where $\rho$ is the
proper density and $U^{\mu}$ is the four velocity), is overlayed in
white in order to define the location of the "funnel wall jet," as
was done in \textbf{HK}. The funnel wall jet is a shear layer
between the accretion disk corona and the Poynting jet. It is a
collimated sub-relativistic flow that transports most of the mass
outflow in the jetted system. In \textbf{HK}, it was shown to be
driven by the total pressure (gas plus magnetic) gradient in the
corona that is oblique to the funnel wall boundary. The gas in this
region is constrained from being pushed into the funnel by the
centrifugal barrier. The component of pressure gradient that is
parallel to the centrifugal barrier forces the flow to be squeezed
outward as a shear layer.
\begin{figure}
\hspace{-0.7cm}
\includegraphics[scale=0.25]{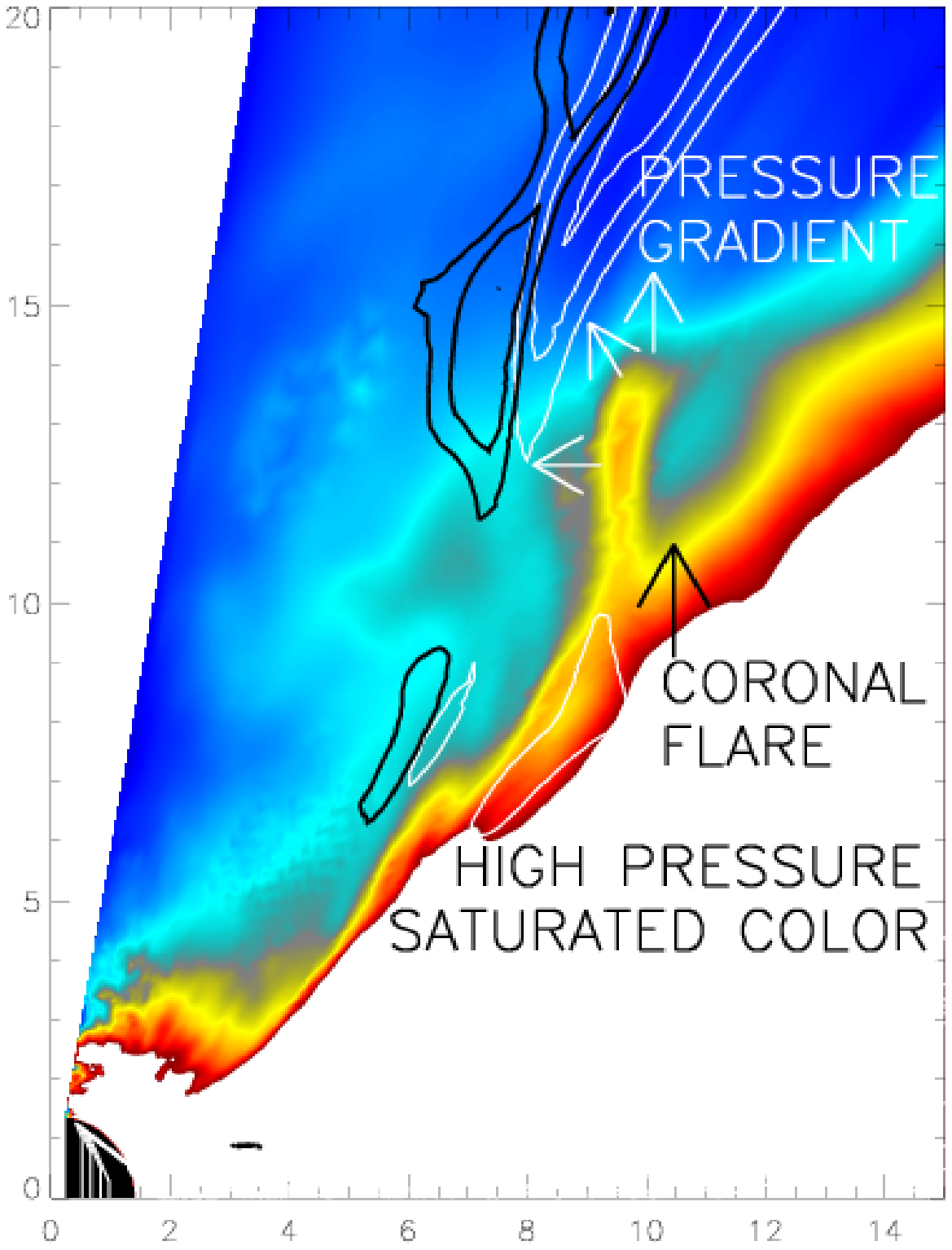}
 \hspace{-2.6cm}
\includegraphics[scale=0.25]{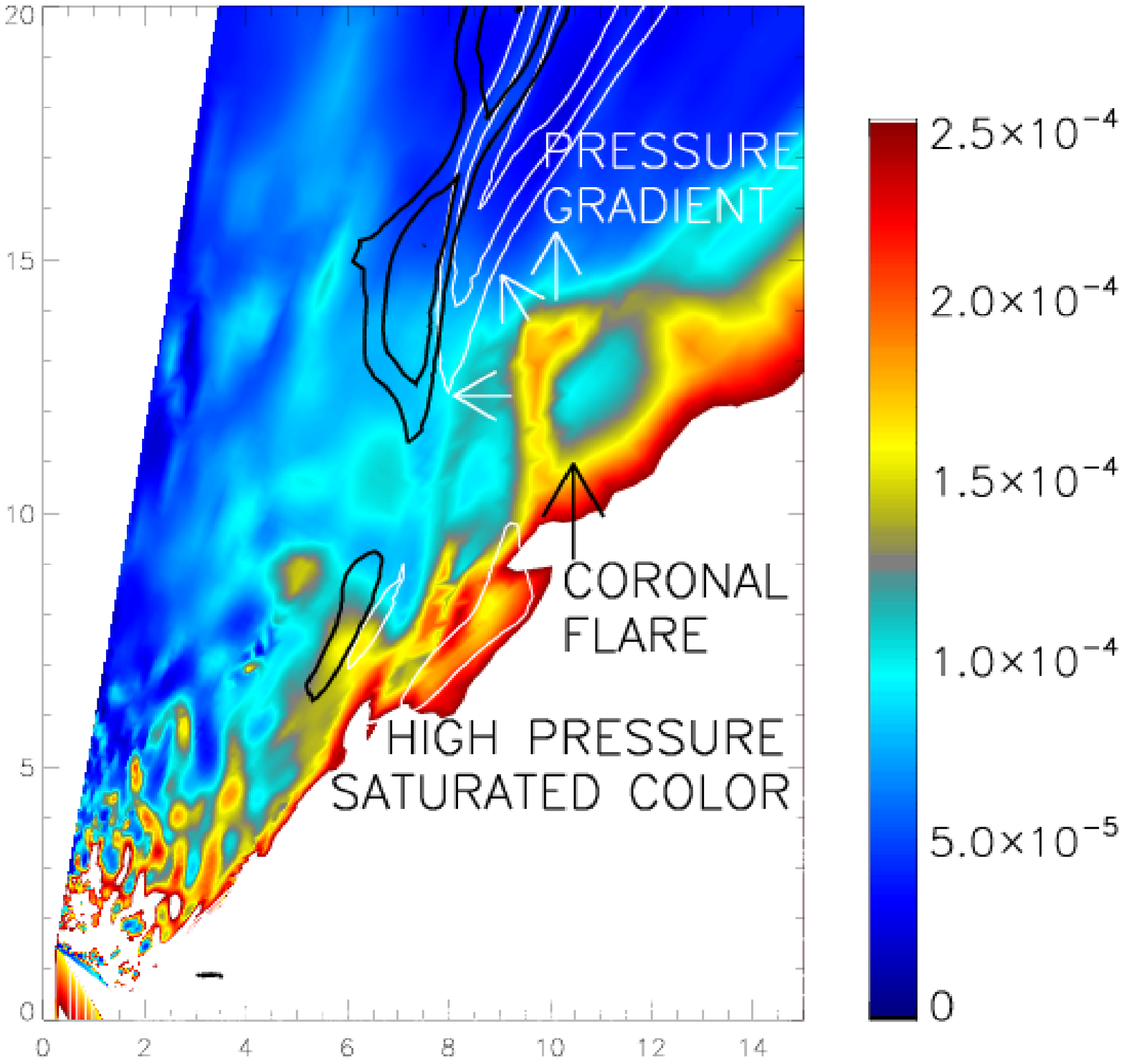}
 \caption{A closeup of the coronal piston at t =10000 M is depicted in these plots of the total pressure density
 (gas plus magnetic), expressed in terms of the gas pressure $P_{g}$ and the Faraday tensor as $ \sqrt{-g}(P_{g} +
 F^{\mu\nu}F_{\mu\nu}/16 \pi)$. The left frame is averaged over $\phi$ and the right frame is at $\phi=49.2^{\circ}$. The color bar is the strength of the total pressure
 in code units. High pressure regions of the corona are saturated and appear white. The force associated with the pressure gradient is indicated by the
 white arrows. The overlayed white contours represent, $P_{r}$, the radial
momentum flux as in figure 2 and the black contours represent
$S^{r}$.}
\end{figure}
\par In figure 2, there is an almost one to one correspondence,
between locations where $P_{r}$ of the funnel wall jet increases and
sites where $S^{r}$ increases at the funnel wall boundary in the
EHM. It is the coronal pressure and not the outgoing Poynting jet
(OPJ, hereafter) that drives the funnel wall jet. Figure 3 shows a
strong flare in the total pressure (gas plus magnetic) at t =10000
M. The pressure gradient seems to provide the accelerating force
that drives $P_{r}$. The flare appears to be a high pressure loop
emerging from the corona, as evidenced by the right hand frame of
figure 3. The loop location and topology are inconsistent with the
high pressure feature being injected into the corona from the funnel
interior. The magnetic pressure in the corona actually exceeds the
magnetic pressure in the EHM at these intermediate radii, $ 10M < r
< 30 M $, precisely the region where the flares tend to occur. Table
4 of \textbf{HK} indicates that $\overline{S_{in}} \equiv \int_{r =
r_{in}} S^{r}\, d\theta d\phi dt \equiv \int S^{r}_{in}\, d \theta =
(2.79/4.26)P_{jet}$, where $P_{jet}$ is the total energy transported
to $r=120 M$ by the funnel wall jet. Table 4 also indicates that $
\int_{r = 120 M} S^{r}\, d\theta d\phi dt $ = (1.46/2.79) $
\overline{S_{in}}$. Therefore, by conservation of energy, a putative
B-Z effect can provide at most $ (2.79 - 1.46)/4.26 = 0.31$ of
$P_{jet}$, and is too feeble to drive the funnel wall jet.
\par Most of $\overline{S_{in}}$ is directed into
the equatorial accretion flow and never reaches the plasma in the
OPJ. To see this, we summarize figure 5 of \textbf{HK}. There are
broad relative maxima in $S^{r}_{in}$ at $\theta < 41.3^{\circ}$
($\theta > 137^{\circ}$), but $S^{r}_{in}$ decreases rapidly with $
\theta $ and $S^{r}_{in} < 0 $ when $41.3^{\circ}< \theta <
54.5^{\circ}$ ($120.4^{\circ}<\theta < 137^{\circ}$). The two
strongest maxima in $S^{r}_{in}$ are in the equatorial accretion
flow at $\theta\approx 80^{\circ}$ ($\theta \approx 100^{\circ}$)
and this energy never reaches the OPJ. Therefore, it is unclear how
the region $41.3^{\circ} < \theta < 137^{\circ}$ is related to the
B-Z mechanism. In order to understand the global energy budget, we
want to know how much $\overline{S_{in}}$ reaches the OPJ. Based on
the available data, we can generate an upper bound. First, we define
the OPJ as in \citet{hir04} by the magnetic dominance condition,
defined in terms of the Faraday field strength tensor as $
F^{\mu\nu}F_{\mu\nu}/(16 \pi \rho h)>1$ ($h$ is the enthalpy per
unit mass) and  $U^{r}>0$. Virtually all the $S^{r}$ that reaches $r
= 120M $ in the available KDH time slices flows in the OPJ as
defined above. Inspecting our three time slices, the OPJ initiates
at $r_{min} < 3M $, with an angular extreme near the base given by
$\theta_{max}\approx 55^{\circ}$ ($\theta_{min} \approx
125^{\circ}$), in agreement with the time averaged data in figure 11
of \citet{haw06}. The OPJ is collimated, i.e., $d\theta_{max}/dr <
0$ for $\theta < 90^{\circ}$ ($d\theta_{min}/dr > 0$ for $\theta >
90^{\circ}$). In each of the time slices, we can compare $\int S^{r}
d\theta d\phi $ near $r_{in}$ above the equatorial accretion flow to
$\int S^{r} d\theta d\phi $ in the OPJ to assess how much
$\overline{S_{in}}$ reaches the OPJ. The simulations have
significant fluctuations. Averaging over r smooths out the spatial
fluctuations and is a more reliable diagnostic than computing fluxes
at a single radius. Unfortunately, there is not enough data to
perform a meaningful time average. In each time slice we find
$\frac{1}{0.597M}\int^{2M}_{r_{in}} \, dr\int^{\theta =
70^{\circ}}_{\theta = 8.1^{\circ}} S^{r}\, d\theta d\phi
> \frac{1}{4M}\int^{9M}_{5M} \, dr \int^{\theta =\theta_{max}}_{\theta =8.1^{\circ}} S^{r}\, d\theta d\phi
$; $\frac{1}{0.597M}\int^{2M}_{r_{in}} \, dr\int^{\theta
=171.9^{\circ}}_{\theta =110^{\circ}} S^{r}\, d\theta d\phi >
\frac{1}{4M}\int^{9M}_{5M} \, dr \int^{\theta
=171.9^{\circ}}_{\theta = \theta_{min}} S^{r}\, d\theta d\phi$. The
6 inequalities tend to indicate that $\overline{S_{in}}$ at $\theta
< 70^{\circ}$ ($\theta
> 110^{\circ}$) is larger (the average excess is $20 \%$) than the
total $S^{r}$ flowing through the OPJ at $r\approx 7M$, $\theta <
50^{\circ}$ ($\theta > 130^{\circ}$) and some of the
$\overline{S_{in}}$ at $\theta < 70^{\circ}$ is absorbed by the
accretion disk and corona at $r \leq 7M$, $\theta > 50^{\circ}$
($\theta < 130^{\circ}$). This excess is apparently large enough to
persist in spite of significant temporal variations. Conversely, the
probability of the hypothesis that random temporal fluctuations have
conspired to mask what is actually a conserved energy flow from
$r_{in}$, $\theta < 70^{\circ}$ ($\theta
> 110^{\circ}$) to the OPJ at $r \approx 7M$ is rejected at the (0.5 per hemisphere)
$1 - (0.5)^{6}=0.9844$ significance level. Integrating the area
under the plot of $S^{r}_{in}$ in figure 5 of \textbf{HK}, one finds
that only 39\% of the total $\overline{S_{in}}$ originates in the
angular range $8.1^{\circ} < \theta < 70^{\circ}$, $110^{\circ} <
\theta < 171.9^{\circ}$. Thus, the data tends to show that
$0.39\overline{S_{in}}$ is a loose upper limit to the maximum amount
of $\overline{S_{in}}$ that can reach the OPJ. Applying this to the
results in table 4 of \textbf{HK} implies that $> [1.46 -
(0.39)(2.79)]/1.46 = 0.26 $ of the $S^{r}$ in the OPJ at $r =120M$
is created at $r > r_{in}$ during the course of the simulation.
Furthermore, integrating over the funnel cross section at both
$r_{in}$ and across the flares, $[\int^{\theta =35^{\circ}}_{\theta
=8.1^{\circ}} S^{r}\, d\theta d\phi]_{\mathrm{flare}}>
1.5(1/0.597M)\int^{2M}_{r_{in}} \, dr\int^{\theta
=70^{\circ}}_{\theta =8.1^{\circ}} S^{r}\, d\theta d\phi$ in the
last two time steps (there is a weak EDJ at t =9820 M that skews our
results, so it is omitted from this analysis). If one includes the
mechanical energy flux as well, there is consistently more than
twice the total energy flux in the strong flares then there is total
energy flux through $r_{in}$, $\theta < 70^{\circ}$. This suggests
that coronal injection sites are sufficiently strong to make up this
apparent $>26\%$ deficit in $S^{r}$.
\par One might be concerned that there are only
$\approx 8$ -10 angular zones between the coronal piston and the
funnel and this leads to significant numerical diffusion. From a
numerics point of view this is much more of a concern than from a
physical point of view. The MHD code is just a simple approximation
to any real turbulent plasma state. The turbulent corona is likely
to have an anomalous resistivity and diffusion should occur
\cite{som00,tre01}. The rate of diffusion cannot be determined by
this simulation. However, the qualitative idea that a strong flare
in coronal energy can in principle reach the funnel interior is
strongly indicated.
\section{Discussion}The philosophy of this paper is not that the simulations of idealized magnetized tori gives us a direct
picture that can be applied to AGN central engines. They are treated
only as virtual laboratories to see what effects might
self-consistently occur near a magnetized black hole. These
simulations have led us to a new realization, the boundaries of the
EHM should be dynamic and are not likely to be passive boundary
surfaces for the magnetic field. It was shown that electrodynamic
energy flux can arise in the EHM as a result of sources radiating
energy from the lateral boundaries. Even if the EHM can be construed
as "force-free," the dynamics of the lateral boundaries are
determined by strong inertial forces that should make them strong
MHD pistons. This circumstance was not anticipated in theoretical
treatments of electrodynamic jets in the EHM \cite{blz77,phi83}. The
fact that electromagnetic energy can come into the EHM from the side
goes right to the heart of the assumptions in the B-Z solution. The
B-Z solution is the perfect MHD solution in which energy
conservation reduces to Poynting flux conservation from the horizon
to a relativistic wind at asymptotic infinity \cite{phi83}. From
this condition, the parameters of the field are uniquely determined
for a given poloidal field distribution, in particular the field
line angular velocity, $\Omega_{F}$, and the total electromagnetic
energy output from the black hole, $\int S^{r}\, d\theta d\phi$. If
there are strong sources of Poynting flux along the lateral walls of
the EHM, the spacetime near the event horizon can not adjust the
system to enforce the B-Z field parameters within the EHM. This is a
direct consequence of the fact that the plasma near the event
horizon in the EHM can not effectively react back on the outgoing
wind or jet and modify its electromagnetic properties because of the
gravitational redshifting of the MHD characteristics
\cite{pun01,pun04}. The plasma near the horizon in the EHM will
passively accept any field parameters imposed by the EDJ and the
accretion disk corona \citep{pun01}. As such, in a general
astrophysical context, the basic parameters such as $\Omega_{F}$ and
$\int S^{r}\, d\theta d\phi$ are indeterminant. Of course, this does
not preclude the possibility of an MHD numerical system evolving
towards B-Z, if the numerical problem is properly constructed
\cite{kom05}. From a physical point of view, the strong forces that
are responsible for compressing the flux down to the horizon still
reside in the "funnel walls," rendering the lateral boundaries as
strong MHD pistons. In a realistic astrophysical setting, inertial
forces in the lateral boundaries are likely to play an important
role, or even a dominant role, in the determination of the jet power
from the EHM.

\section*{Acknowledgments} I would like to
thank Jean-Pierre DeVilliers, Julian Krolik and John Hawley for
sharing their data and expertise.

\end{document}